\documentclass[twocolum n, superscriptaddress, prb, aps]{revtex4-1}
\usepackage{amsmath}
\usepackage{amssymb}
\usepackage{amsfonts}
\usepackage{bm}
\usepackage{graphicx}
\usepackage{epsfig}
\usepackage{dcolumn}
\usepackage{txfonts}
\usepackage{makeidx}
\usepackage{color}
\usepackage{mathtools}
\usepackage{threeparttable}
\usepackage[colorlinks,linkcolor=red,anchorcolor=black,citecolor=green]{hyperref}

\begin{document}
\title{Cannikin's Law in Tensor Modeling:\\
A Rank Study for Entanglement and Separability in Tensor Complexity and Model Capacity}

\author{Tong Yang} 
\affiliation{Department of Physics, Boston College, Chestnut Hill, Massachusetts, USA}
\affiliation{Machine Learning Center of Excellence, J.P. Morgan, Kowloon, Hong Kong}
\affiliation{Department of Computer Science, Brandeis University, Waltham, Massachusetts, USA}
\affiliation{Center for Polymer Studies, Boston University, Boston, Massachusetts, USA}

\begin{abstract}
	This study clarifies the proper criteria to assess the modeling capacity of a general tensor model.
	The work analyze the problem based on the study of tensor ranks, which is not a well-defined quantity for higher order tensors.
	To process, the author introduces the separability issue to discuss the Cannikin's law of tensor modeling.
	Interestingly, a connection between entanglement studied in information theory and tensor analysis is established, shedding new light on the theoretical understanding for modeling capacity problems.
\end{abstract}

\maketitle

\section{Introduction}


Along with the significant growth of computing power, complicated models becomes available for problems with large degrees of freedom, which, in recent years, has been further popularized along with the progress in deep learning learning research.
People are generally interested in analyzing high-order tensors of large scales, and discussing their capability to capture complex relations.
Efficient tensor models are desired to solve real life problems with fewer adaptive parameters.

Despite the proliferation of related work in  theoretical analysis on tensors, there actually has been a long-time mis-understanding about the modeling power of tensors.
Given a tensor, there are two related while completely different quantities: tensor complexity, and its model capacity.
While the latter is the concern of most research, studies frequently take the problem of the former one to analyze.
Briefly speaking, tensors with large complexity is not guaranteed to be a model with sufficient capacity.
This confusion motivates the current work, which delivers a comparison between these two perspective, including problem setup, theoretical analysis, and related techniques.

More specifically, in the field of tensor analysis, in general, people are interested in low rank efficient representations of high order tensors. There are mainly two different sets of problems:
\begin{enumerate}
\item 
One does not tries to approximate a specific tensor, but is more interested in finding an efficient model space.
\item 
One aims at approximating a specific tensor, given exact information about tensor entries.
\end{enumerate}
For the first set of problems, one focus more on tensor model structures, rather than an algorithm to find the optimal approximation. Therefore, the model capacity would be the foremost concern. In the second set of problems, the ultimate goal is to find a best (or sub-optimal) estimation with direct information on entries, therefore one would compose an explicit algorithm for higher order tensor decompositions, which produces a resulting model structure (e.g. Tucker-decomposition producing the Tucker-format, and sequential SVD producing the tensor-train format).

Due to the above difference, theoretic analysis associated with the two set of problems also differs a lot. On the one hand, to capture the model capacity, one popular choice is the so-called canonical polyadic (CP) rank: a higher CP-rank is usually regarded as a sign of higher capacity. On the other hand, the task of the tensor approximation requires that the model class forms a closed set, which rules out most model structures containing loops, leaving tree tensor models more popular among the community that can be optimized with generalized versions of SVD.

The major goal of this work is to
clarify a proper analysis scheme for investigating tensor model capacity, and therefore to provide further insight for designing efficient models.
However, we would start by arguing that the CP-rank is not a proper language for this purpose, as tensor complexity and model capacity are conceptually different. Instead, we apply the idea of truncating small weight Schmidt components, and clarify assumptions of "separability" implied by any low order tensor model structure. 

To achieve this, we start by introducing the generalized Schmidt decomposition and finite rank truncation, along with some popular algorithms which help to find a quasi-optimal solution. 
Then we introduce the definition of CP-rank as a generalized version of matrix rank, which can be used for tensor complexity analysis. We continue by clarifying the difference between tensor complexity and model capacity, and provides a more natural capacity measure: the \emph{separability scaling behavior} (SSB). With this measure, different existing tensor model structures are compared and further insights could be derived for model design in black-box modeling tasks.

\section{Generalized Schmidt Decomposition and Relevant Algorithms}

The problem of multivariate function tensorization (MFT) and tensor approximation (TA) problems are similar to each other. For MFT, one is usually given a function defined on $\mathbf{\mathcal{I}}\subset\mathbb{R}^L$: $f(x_1, x_2, \cdots, x_L)$, and attempt to decompose it into a product of single variate orthonormal basis functions. The TA problem, on the other hand, aims at decomposing a high order tensor $\mathcal{A}_{s_1s_2\cdots s_L}$ into a product of low order tensors, where $s_i\in [1, D]$. 
In TA problems, one needs $D^L$ number of entries to specify a tensor; while in MFT, one instead requires $P^L$, where $P$ represents the number of single variate orthonormal basis functions, and in general could be $\infty$. 

In general, as the complexity (for both storage and computation) increase exponentially when "order" increase, one is interested in a "low order decomposition" to approximate a tensor/function, where the term "order" means number of modes/variables in tensors/functions.

\subsection{General Schmidt Decomposition}
In both problems, the Schmidt decomposition plays the key role, which reveals the interplay (entanglement) between different modes/variables. More precisely, consider any bipartition of modes/variables which leads to a matricization of the original tensor/function:
\begin{align}
    \mathcal{A}_{s_1s_2\cdots s_L} &= A_{s_a, s_b}; \nonumber \\
    f(x_1, x_2, \cdots, x_L) &= f(x_a, x_b).
\end{align}
One could then apply a Schmidt decomposition on the two separated parts, by finding the left and right singular vectors/functions defined independently on the two parts:
\begin{align}
    \quad u^{\alpha}_{s_a}\;\; &\quad ,\quad\;\;\: v^{\alpha}_{s_b} \nonumber \\
    \psi^{\alpha}(x_a)  &\quad ,\quad \phi^{\alpha}(x_b) \qquad \forall\alpha\in[1, R]
\end{align}
where $u^{\alpha}$ and $v^{\alpha}$ are left and right singular vectors of the matrix $A$, whose indices are labeled as $s_a$ and $s_b$ running from 1 to $d_a = D^{l_a}$ and $d_b = D^{l_b}$,  respectively; while $\psi^{\alpha}(x_a)$ and $\phi^{\alpha}(x_b)$ are left and right singular functions of the bi-variable function $f(x_a, x_b)$, whose variables are labeled as $x_a$ and $x_b$ respectively (taking values in a bounded region). The index $\alpha$ labels different singular vectors/functions, where, in total, $R$ of them exist: $R$ is bounded by $\min(d_a, d_b)$ in matrix case used for TA and in general reaches $\infty$ in function case used for MFT. The resulting Schmidt decomposition then follows:
\begin{align}\label{schmidt}
    A_{s_a, s_b} &= \sum_{\alpha=1}^{R} \sqrt{\lambda_{\alpha}}\cdot u^{\alpha}_{s_a} \cdot v^{\alpha}_{s_b}; \nonumber \\
    f(x_a, x_b) &= \sum_{\alpha=1}^{R} \sqrt{\lambda_{\alpha}}\cdot \psi^{\alpha}(x_a) \cdot \phi^{\alpha}(x_b),
\end{align}
where $\lambda_{\alpha}$ is called singular values, which is assumed to be arranged in a descending order. 

An important feature of this decomposition is that each component are orthogonal to each other, and therefore expands in an independent dimension (subspace). With a proper defined inner product (dot product for vectors, and $L_2$-integral on a bounded region for functions), the distance between two arbitrary tensors/functions can then be expressed with merely the coefficients $\lambda_{\alpha}$'s, which provides a necessary criteria to discuss tensor/function approximation.

Going back to Eq\eqref{schmidt}, instead of using all $R$ components which leads to an exact expression, one could use a truncated expression up to $r$ components as an approximation, regarding the fact that coefficients (singular values) are in a descending order. the resulting $L_2$-error in both case is then:
\begin{align}\label{error_general}
    \epsilon = \sum_{\alpha=r+1}^{R} \lambda_{\alpha}.
\end{align}
One is usually interested in a low rank approximation, which corresponds to a small number $r$. To achieve an efficient low rank approximation, the descending sequence $\{\lambda_{\alpha}\}$ must vanish fast enough. We could consider the two extreme cases:
\begin{itemize}
    \item in the worst scenario, all $\lambda_{\alpha}$'s are equal without vanishing, which would lead to a worst approximation (corresponding to the "maximally entangled" case);
    \item in the best scenario, all but the first $\lambda_{\alpha}$ are zero, meaning the tensor/function can be written as a product of two tensors/functions defined in orthogonal spaces (corresponding to the "disentangled" case).
\end{itemize}
In more general cases, given an error acceptance threshold $\epsilon$, the minimum number $r$ of components kept to achieve the error threshold therefore suggests the difficulty of the approximation of a tensor/function: the larger $r$ is required, the more difficult the tensor/function approximation is.

The above discussion implies that the set of singular values $\{\lambda_{\alpha}\}$ (also called entanglement spectrum in physics) actually captures \textbf{the "separability" of two parts given a bipartition.} The information contained in the spectrum could be extracted in multiple levels:
\begin{itemize}
    \item the number of non-zero $\lambda_{\alpha}$'s, i.e. matrix rank, which also relates to the zeroth order R\'enyi entropy;
    \item the distribution of $\{\lambda_{\alpha}\}$, which can be further captured by the $L_n$-distance from an uniform distribution (all $\lambda_{\alpha}$'s are equal), related to the $n$-th order R\'enyi entropy.
\end{itemize}
This motivates one to use the entanglement entropy to categorize different problems.

\textcolor{black}{\subsection{The (quasi) Optimal Approximation of Low Order Decomposition}}

\textcolor{black}{The above introduced Schmidt decomposition in general setups not only provides a way to analyze the error in this specific approach, but also implies a method to achieve a quasi-optimal (if not the best) approximation with a low order decomposition. The capability for achieving the "quasi-optimal" approximation is rooted in two facts: the orthogonality of different components, and the exact error expression in Eq\eqref{error_general}. Indeed, \textbf{given rank $r$ for a bipartition approximation (express a tensor/function using two lower order tensors/functions), the best approximation minimizing the Frobenius norm criteria is achieved by truncating the subspace expanded by components indexed higher than $r$ in the Schmidt decomposition.}}

\textcolor{black}{To put it more systematically, low order decompositions involve connecting (contracting) pieces of low order components, i.e. lower order tensors in TA and functions with fewer variables in MFT. There is no algorithm that could find the best, or even the quasi-optimal, approximation (global optima) for generic tensor models. For example, both the CP-decomposition and tensor networks containing closed cycles consist of a tensor set that is not closed, which renders the problem of finding a best approximation ill-posed. However, for tree tensor networks, there is indeed a general recipe to find at least the quasi-optimal approximating tensor: a sequential Schmidt Decomposition (SSD), which corresponds to the high-order singular value decomposition (HOSVD) in tensor analysis field.}

\textcolor{black}{There are different versions of SSD that are associated with different tensor structures. \textbf{Generally speaking, a SSD consists of a sequence of Schmidt decompositions acting on different modes/variables. Each Schmidt decomposition slices out a lower order components that connects with the rest modes/variables through a single-leg tensor contraction.}}

\textcolor{black}{For example, the quasi-optimal approximation of Tensor-Train models can be achieved by slicing one mode/variable each time through a Schmidt decomposition; the quasi-optimal approximation of Tucker-format tensor models can be achieved by applying Schmidt decomposition on each single mode/variable individually; The Hierarchical-Tucker models can be quasi-optimally approximated through a root-to-leaves sequence of Schmidt decompositions.}

\textcolor{black}{The error analysis of algorithms differ in the tensor and function cases: in MFT problems, error could be bounded by certain constant (usually as a function of number of variables and tensor ranks); while in TA problems, given tensor ranks, error could only be bounded by the minimum error of the model structure itself multiplied by a constant (usually as a function of tensor orders). Briefly speaking, this is due to the fact that in MFT, there are usually certain extra assumptions on the smoothness of the function, while, in TA, for high order tensors with finite number of entries, there are no such constraints.}

\section{Tensor-Complexity Analysis: Canonical Polyadic Rank}

The above discussed TA problem describes the case when direct information (entries) is given about the target tensor, one is interested in finding a low order tensor approximation. The general procedure contains a sequence of SVDs that subsequently slice off mode clusters, and the final approximation are composed by a series of virtual index contractions.

Regardless of the general recipe, another important question is the complexity of a tensor, which renders the difficulty of approximation in practice. For higher order tensors, the most popular criteria describing tensor complexity is the canonical polyadic rank, which can be deemed as a generalization of matrix rank for higher order cases. Basically, a tensor with larger canonical polyadic rank is associated with a higher complexity.
In this section, we would introduce this widely used concept.

\subsection{Canonical Polyadic Rank}
The Canonical Polyadic (CP) rank can be regarded as a generalized version of matrix rank in the case of higher order ($>2$) tensors. We could discuss matrix rank using Schmidt decomposition form:
\begin{align}\label{mat_schmidt}
    A_{s_a, s_b} &= \sum_{\alpha=1}^{R} \sqrt{\lambda_{\alpha}}\cdot u^{\alpha}_{s_a} \cdot v^{\alpha}_{s_b}, \qquad s_{a, b}\in [1, d_{a,b}].
\end{align}
The number $R$ is the rank of the matrix $A$. As mentioned before, the rank captures the separability of two parts from a bipartition of modes. For higher order tensors, the CP rank is defined in a similar way, where a bipartition is replaced by a multi-partition:
\begin{align}\label{cp}
    \mathcal{A}_{s_1s_2\cdots s_L} &= \sum_{\alpha=1}^{R} \sqrt{\lambda_{\alpha}}\cdot v^{\alpha}_{s_1} v^{\alpha}_{s_2}\cdots v^{\alpha}_{s_L},
\end{align}
and $R$ is then called the CP-rank of tensor $\mathcal{A}$. As now the system is partitioned in multiple modes, it is not clear how to define the separability issue in this form now. In fact, \textbf{CP rank is equivalent to (up to a exp/log function) the Schmidt measure for multipartite entanglement.}
In quantum information field, it is well know that the Schmidt measure cannot distinguish between truly multipartite entanglement and bipartite entanglement.

\subsection{Upper Bound Nature of CP Rank}
Further more, we would like to relate the CP rank and matrix rank. As used above, a popular trick for higher order tensor is matricization through a mode bipartition, i.e.
\begin{align}
    \mathcal{A}_{s_1s_2\cdots s_L} \quad \longrightarrow \quad A_{s_a, s_b}.
\end{align}
One could easily prove that \textbf{CP rank $R$ is an  upper bound of matrix ranks for all possible matricizations.} 

    Basically, we prove $rank\big[A^{(a,b)}\big]<R$ for any matricization with mode bipartition $(s_a, s_b)$, where $A^{(a,b)}$ is the matricization partitioning the mode $(s_1, s_2, \cdots s_L)$ into $(s_a, s_b)$. For each component in a CP decomposition, the matrix rank is 1, since it is a direct product state and purely separable. Then due to the linear nature of the matricization operation:
    \begin{align}
        A^{(a,b)} :=& \bigg[\sum_{\alpha=1}^{R} \sqrt{\lambda_{\alpha}}\cdot v^{\alpha}_{s_1} v^{\alpha}_{s_2}\cdots v^{\alpha}_{s_L}\bigg]^{(a,b)} \nonumber \\
        =& \sum_{\alpha=1}^{R} \big[\sqrt{\lambda_{\alpha}}\cdot v^{\alpha}_{s_1} v^{\alpha}_{s_2}\cdots v^{\alpha}_{s_L}\big]^{(a,b)}, 
    \end{align}
    we therefore have:
    \begin{align}
        rank\big[A^{(a,b)}\big] &= rank\bigg[\sum_{\alpha=1}^{R} \big[\sqrt{\lambda_{\alpha}}\cdot v^{\alpha}_{s_1} v^{\alpha}_{s_2}\cdots v^{\alpha}_{s_L}\big]^{(a,b)}\bigg] \nonumber \\
        &\leq \sum_{\alpha=1}^{R} rank\bigg[\big[\sqrt{\lambda_{\alpha}}\cdot v^{\alpha}_{s_1} v^{\alpha}_{s_2}\cdots v^{\alpha}_{s_L}\big]^{(a,b)}\bigg]\nonumber \\
        &= R
    \end{align}
    Therefore the CP rank $R$ is an upper bound of matrix rank among all possible matricizations.
    
    

Therefore, although CP-rank cannot be directly defined as "separability", it at least captures the upper bound of all possible matrix ranks. In other words, \textbf{the CP-rank describes the separability of the most inseparable mode bipartition.} In this sense, it can indeed serve as a tensor complexity measure, although more details of interactions (entanglements) among different modes are absent.

\section{Tensor Model Capacity Analysis: Coordinate Separability}

Above we introduced two important concepts: 
\begin{itemize}
    \item The sequential Schmidt decomposition provides both a class of algorithms to construct lower order (tree) tensor approximation for high order tensors and associated error analysis;

    \item The CP-rank describes the complexity of a tensor, which is usually the target tensor to be approximated. 
\end{itemize}
With the help of the above discussion, we now study the problem that is more crucial in practice: constructing a high-capacity model. Most Deep Learning researches are more related to this category: since deep neural network models are usually optimized using gradient based methods, the goal of changing model structure is then not to design a model that could be cheaply optimized by a novel algorithm, but purely to find a novel model which, itself, could capture desired features and dynamics of the downstream task. A meaningful criteria to evaluate the model capacity is hence desired.

In many previous studies, the same CP-rank has been used to describe model capacity. We would like to argue that this may not be the best criteria for model analysis, based on which, we would like to return to the separability property, and describe the model capacity using the scaling behavior of bipartition matrix ranks.

\subsection{Difference between Model Capacity and Tensor Complexity}
Firstly, we would like to clarify the difference between the complexity of a tensor and the capacity of a model. 

As we discussed earlier, the technique using lower order pieces to construct higher order tensors relies on the separability issue. In the case where all modes a completely separable, only a linear number of basis vectors (rank-1 tensor) are required; in the case where any two parts are inseparable, it is quite difficult to construct a lower order representation. Therefore, both for tensor complexity and model capacity, we would discuss the separability issue.

On the one hand, the tensor complexity implies how difficult it is to describe a tensor. \textbf{The most difficulty would appear in the bipartition with a highest matrix rank, which in the lower order representation would result into a contraction with higher virtual dimension.}

On the other hand, the model capacity should be evaluated by considering "weakness" in the structure. Given a tensor model, one could also consider different mode bipartitions. In this scenario, however, the bipartitions associated with lower matrix rank should be concerned: \textbf{by applying the corresponding model, one has \emph{assumed} a strong separability on these mode bipartitions.}

As we proved above, the CP-rank provides an upper bound for bipartition matrix rank, and therefore could serve as a primitive and basic description for tensor complexity. But for model capacity, the CP rank may not be a proper criteria, as it does not contain the information of mode bipartitions with lower matrix ranks.

\subsection{Black-Box Tensor Modeling Problems}

Most generally, we could answer a more straightforward question: suppose a separability assumption is made on a $L$-order target tensor $\mathcal{A}_{s_1s_2\cdots s_L}$, what is the minimum universal virtual-leg dimension $R$ that could guarantee the target tensor being well-approximated, when using different model structures?

A black-box modeling procedure could be performed to approximate a target tensor with certain separability assumption (or say, information). Given a tensor model structure $\mathcal{M}$, different modes-bipartitions could capture interplay (entanglement) with different complexity, However, note as a black-box modeling, one in general cannot arrange different modes/variables in a way that the target tensor complexity and the model complexity match each other. Instead, to guarantee a solution to be found, the following relation should hold:
\textcolor{black}{
\begin{align}\label{minmax}
    \max_{a\in \mathcal{P}_m}{rank\big[A^{(a, \Bar{a})}\big]} \leq 
    \min_{a'\in \mathcal{P}_m}{rank\big[A^{(a', \Bar{a}')}_{\mathcal{M}}\big]}, \quad \forall m\in [1, L/2],
\end{align}
where $A^{(a, \Bar{a})}$ is the matricization of the target tensor $\mathcal{A}$ associated with the modes-partition $(s_1, s_2, \cdots, s_L) = s_a\cup s_{\Bar{a}}$, and $A^{(a', \Bar{a}')}_{\mathcal{M}}$ is the matricization of the model tensor $\mathcal{A}_{\mathcal{M}}$ associated with the modes-partition $(s_1, s_2, \cdots, s_L) = s_{a'}\cup s_{\Bar{a}'}$. And $\mathcal{P}_m$ is the set of modes-bipartitions where the smaller part contains $m$ modes.}

With the above discussion, it becomes quite clear that the model capacity, which is captured on the right hand side, is related to the lower bound of matricization ranks.
We term the above relation in Eq.\eqref{minmax} as the \emph{Cannikin's law of tensor modeling}.

\subsection{Strong Separability Assumption in Popular Tensor models}

From the above analysis we are aware that, given any model structure, to analyze the model capacity, one should put particular concern on the mode bipartitions associated with lower matrix ranks. Now we analyze some popular tensor models as a further demonstration.

\subsubsection{Tensor-Train Model}

The Tensor-Train models (TT)~\cite{yu} construct higher order tensors in the following form:
\begin{align}
    \mathcal{A}_{s_1s_2\cdots s_L} = \sum^{r}_{\{\alpha_i\}} M_{s_1}^{\alpha_1} M_{s_2}^{\alpha_1,\alpha_2}\cdots M^{\alpha_{L-1}}_{s_L},
\end{align}
where each $M$ is a order-3 tensor (except the boundary two which are of order 2). For simplicity while w.l.o.g, we consider the case where are virtual bonds have the same dimension $r$.

Firstly, consider any bipartition separating the sequence $(s_1s_2\cdots s_L)$ with one cut, i.e. $(s_1\cdots s_m)\cup(s_{m+1}\cdots s_L), \forall m\in[1, L-1]$, where we label two disjoint sets as $s_a = (s_1\cdots s_m)$ and $s_{\Bar{a}} = (s_{m+1}\cdots s_L)$. The resulting matrix rank can be calculated as:
\begin{align}
    &\; rank\big[A^{(a, \Bar{a})}\big] \nonumber \\
    = &\; rank\bigg[\sum^r_{\alpha_m} \bigg(\sum_{\{\alpha_i\}}M_{s_1}^{\alpha_1}\cdots M_{s_m}^{\alpha_{m-1},\alpha_m}\bigg)\cdot \bigg(\sum_{\{\alpha_j\}}M_{s_{m+1}}^{\alpha_{m},\alpha_{m+1}}\cdots M^{\alpha_{L-1}}_{s_L}\bigg)\bigg] \nonumber \\
    = &\; rank\bigg[\sum^r_{\alpha_m}U_{s_1\cdots s_m}^{\alpha_m}U_{s_{m+1}\cdots s_L}^{\alpha_m}\bigg] \nonumber \\
    = &\; rank\bigg[U_a \cdot U_{\Bar{a}}\bigg],
\end{align}
where the two matrices $U_a$ and $U_{\Bar{a}}$ are contracted through an inner product of virtual bond $\alpha_m$, which has dimension $r$. We can therefore conclude that such bipartitions produce matrix ranks upper-bounded by $r$. 

More generally, any cut may result the upper bound of matrix ranks increase by $r$. Therefore, the bipartitions associated least lowest possible matrix rank are those created by one single cut.

Now we interpret the meaning of such mode bipartitions. As shown above, in the matrix expression:
\begin{align}
    A^{(a, \Bar{a})}
    = \sum^r_{\alpha_m}U_{s_1\cdots s_m}^{\alpha_m}U_{s_{m+1}\cdots s_L}^{\alpha_m},
\end{align}
the virtual bond contraction only involves $r$ terms. This is equivalent to assume that the interplay between the modes clusters $(s_1\cdots s_m)$ and $(s_{m+1}\cdots s_L)$ can be efficiently captures by only $r$ terms. In the case where $r\rightarrow \infty$, this decomposition could be exact, without any error introduced. In practice, however, we are interested in finite $r$ case, with the hope that the truncation in $r$ would be accepted given an error threshold $\epsilon$.

More precisely, w.l.o.g, we assume $m\leq \frac{L}{2}$ (hence $D^m\leq D^{L/2}$), and each original mode $s_i$ could take $D$ different possible values. The dimension of the space expanded by $(s_1\cdots s_m)$ is therefore $D^m$. The $r$ terms summation would be sufficient to capture the interplay between two clusters in the case where $r\geq D^{m}$; if $D^m\geq r$, which in a high order tensor problem is very possible, then the $r$ terms summation in general would miss certain interplay between modes clusters (i.e., truncated terms in the full summation), and eventually introduce errors. \textbf{In other words, by implementing a TT-model with finite virtual dimension $r$, one assumes each $m$-modes cluster (starting from one end) interacts with the rest modes through $r$ terms summation.}

To restore an arbitrary $L$-order tensor, the required universal virtual dimension of a TT model would then be:
\begin{align}
    R_{TT} = D^{\frac{L}{2}}.
\end{align}


\subsubsection{Hierarchical-Tucker Model}

A $H$-level Hierarchical-Tucker model constructs higher order tensors in the following form:
\begin{align}
    \phi_{\alpha}(h+1, j) = \sum_{\beta_1, \beta_2} \Lambda^{\beta_1, \beta_2}_{\alpha}(h+1,j)
    \cdot\bigg[\phi_{\beta_1}(h, 2j-1)\cdot\phi_{\beta_2}(h, 2j)\bigg]. \nonumber
\end{align}
Each $(h, j_h)$ is a coordinate in the quasi 2-dimensional tree structure, $h\in[0, H]$ representing the layer index in the tree, and $j\in[1, l_h]$ represents the translational index in the layer. 

The above form represents a bi-HT model: each higher layer tensor is obtained from only two lower layer tensors through an order-3 coefficient tensor $\Lambda(h, j_h)$. In this case, one could easily derive that:
\begin{align}
    H = \log_2{L}; \qquad l_h = \frac{L}{2^{h}}.
\end{align}
The zero-th layer mode tensors have the following form:
\begin{align}
    \phi_{s_i}(0, i), \qquad \forall i\in[1, L],
\end{align}
which, together with all order-3 coefficient tensors $\Lambda(h, j_h)$, determines the large $L$-order tensor. The last layer coefficient tensor could be an order-2 tensor: $\Lambda^{\beta_1,\beta_2}(H, 1)$.

Similar to the analysis on TT-models, again we consider all virtual dimensions are the same, i.e. all three indices of $\Lambda$ runs within $[1. r]$ (except the zero-th layer). The bipartitions with only one single cut then correspond to a relatively large truncation. In fact, a cut slicing off $m=2^h$ modes from one end, in general, requires $r=D^m$ virtual dimension. The cut on the last layer requires $r=D^{L/2}$ for an exact representation of an arbitrary $L$-order tensor, which is also the one that may introduce largest  errors when a finite dimension truncation is applied. \textbf{In other words, by implementing a HT-model with finite virtual dimension $r$, one assumes each $2^h$-modes cluster (starting from one end) interacts with the rest modes through $r$ terms summation.}

To restore an arbitrary $L$-order tensor, the required universal virtual dimension of a HT model would then be:
\begin{align}
    R_{HT} = D^{\frac{L}{2}}.
\end{align}

\subsection{Weaker Separability Assumptions}

The above study has clarifies the separability assumptions implied by popular tensor model structures. \textbf{Generally, when constructing higher order tensors from lower order ones through virtual bond contractions with finite bond dimension $r$, the model implicitly assumes the interplay between two modes-clusters could be well-approximated by $r$ terms.} This assumptions becomes more difficult to be satisfied when the contracted two clusters expand a larger space.

We call a separability assumption "\emph{strong}" when the number of summed terms capturing the modes interplay is much smaller than the total dimension of the interplay space, i.e. the \emph{truncatable} space is large. 

In general, a tensor model should prevent stronger separability assumptions, unless the target tensor satisfies certain special conditions (e.g. area law functions in low-energy states of strongly correlated system). The popular tensor models analyzed above, however, when using finite tensor rank $r$. have implicitly added many strong separability assumptions on the cuts that separate two large clusters.

When no specific information is provided about the target tensor, e.g. modeling with deep neural networks, one in general expect the interplay between two clusters becomes more complicated when the cluster dimension increases. Therefore, when the separated two clusters have a higher dimension, the model structure should also involve more summed terms, which, given a universal virtual bond dimension, implies a larger number of contracted bonds.

More specifically, one aims at finding tensor model structures that contain more bonds contractions when separated modes-clusters expand larger spaces. Suppose the smaller cluster in a bipartition involve $m$ modes, then the number of contracted indices $n$ should increase when $m$ increases.

There could be different scaling behaviors of $n$ as a function of $m$, which in general depends on the targeted tensors. The worst scenario corresponds to the case where $n(m)$ is an exponential function; in this case, however, it is impossible to construct low rank tensor models, as any truncation would result into large errors, and we call it as an "irreducible problem". The best scenario may require only a constant number $n$ of contracted indices, which is the assumption made by both TT and HT models; however this is apparently an extremely strong assumption that is difficult to be satisfied. \textbf{One is hence interested in model structures with weaker separability assumptions, such that $n(m)$ is a monotonically increasing function, but not grow exponentially.} We use the term \emph{separability scaling behavior} (SSB) to describe the function form of $n(m)$. The TT and HT models therefore have constant separability scaling, and the irreducible case has an exponential separability scaling.

Two typical behaviors of $n(m)$ are power-law $n\sim m^{\alpha}$, and logarithm $n\sim \log{m}$. 
Given a model structure, if \emph{any} modes bipartitions satisfy \emph{at least}:
\begin{itemize}
    \item an exponential separability scaling, then we call the model an exponential separable mode; 
    \item a power law with an exponent $\alpha$ separability scaling, then we call the model a power-$\alpha$ separable mode; 
    \item a logarithm separability scaling, then we call the model a logarithm separable mode; 
    \item a constant separability scaling, then we call the model a constant separable mode.
\end{itemize}
Among all classes, the exponential separable models are irreducible, i.e. there does not exist a efficient low rank representation.

Importantly, in the above definition, the term \emph{"at least"} implies the fact that there may exist bipartitions associated with more complicated separability scaling behaviors, and emphasizes that in general one should be concerned with the lower bound of all possible scaling behaviors. \textbf{This contrasts the CP rank analysis of tensor complexity, due to the difference we emphasized earlier: the complexity analysis of a target tensor depends on the most complicated interplay among different modes, which corresponds to the upper bound of bipartition matricization ranks; while the capacity analysis of a model structure depends on the strongest assumption made on the "separability" issue, which corresponds to the lower bound of separability scaling.}

\section{Tensor Models with Weak Separability Assumptions}
In this section, we introduce a new tensor model structure that implies a weaker separability assumption, which is easier to be satisfied in practice compared with TT or HT models. The model is called Multiscale Entanglement Renormalization Ansatz (MERA), and belongs to the logarithm separable model category.

\subsection{Multiscale Entanglement Renormalization Ansatz}

MERA is proposed in the field of quantum information, to capture more complicated quantum states beyond an area-law scaling of EE.
We would briefly mention MERA, and more details can be found in \textit{Evenbly and Vidal 2007}\cite{evenbly} , etc. 
The general idea of MERA, different from other TNs, is to use a $(d+1)$-dim TN to represent a $d$-dim system, where the extra dimension in physics represents the flow of the Renormalization Group (RG). It has been noticed before that a MERA structure is quite similar to CNN\cite{yahui}.

There are essentially two major types of tensor blocks in MERA: disentangler tensors and isometry tensors.
A MERA representation of a general high-order tensor follows a hierarchical structure.
Taking the spatial dimension to be 1, and set the original tensor order $L=4$, the MERA representation of the tensor could be written as:
\begin{align}
\mathcal{A}_{s_1s_2s_3s_4} &\simeq \sum_{\{q_i, r_j\}}\tilde{\tilde{V}}_{q_1q_2}\tilde{V}^{q_1}_{r_1r_2}\tilde{V}^{q_2}_{r_3r_4}\hat{V}^{r_1r_2}_{s_4s_1}\hat{V}^{r_3r_4}_{s_2s_3}.
\end{align}
where each $\hat{V}$ is an order-4 tensor, termed as a disentangler, and each $\tilde{V}$ is an order-3 tensor, termed as an isometry tensor.
The top-tensor $\tilde{\tilde{V}}$ is always of order 2, which could be viewed as a coefficient tensor.
For higher orders with larger $L$ value, the construction could be easily generalized hierarchically.
For 1d cases, MERA describes systems whose EE scales with a $log$-correction, which enters in MERA as a result of spatial coarse-grain in RG. 
Hence MERA would be a nice candidate for $log$-correction problems, which is more complicated than area-law ones.

\subsection{Separability Assumption in MERA}

Suppose a universal virtual bond dimension exists, we are interested in the modes-bipartition associated with the least number of bond contractions, which corresponds to the strongest separability assumption implied by the structure. For a mode sequence $(s_1, s_2, \cdots s_L)$, the bipartition at position $m=2^h$ cut \emph{at least} $n(m)\propto \log_2{m}$ virtual bonds (either isometry bonds or reshaped disentangler bonds). As a reshaped disentangler bond (with probability $1$) has virtual dimension $r^2$, below we could focus on the case where all cut bonds are isometry bonds only with virtual dimension $r$.

Again, as $m$ increases, the total dimension of the (smaller) space from a bipartition increases as $D^m$. As, for any form of bipartitions, there are at least $\log_2{m}$ bonds, each of which contracts $r$ virtual dimensions, the model structure guarantees at least $r^{\log_2{m}}$ terms in the summation capturing the interplay between any two parts. If no truncation is allowed, then the virtual dimension $r$ required would be:
\begin{align}
    r(m) = D^{\frac{m}{\log_2{m}}}.
\end{align}
To restore an arbitrary $L$-order tensor, the required universal virtual dimension of a MERA model would then be:
\begin{align}\label{mera_max}
    R_{MERA} = D^{\frac{L}{2(\log_2{L}-1)}}.
\end{align}
Compared with TT and HT models, the required universal virtual dimensions have the following relation:
\begin{align}
    R_{HT} = R_{TT} = \big[R_{MERA}\big]^{\log_2{L}-1}
\end{align}

\subsection{Black-Box Modeling with Given Separability Assumptions}

Now we provides a quantitative analysis for the question raised in earlier sections: the black-box tensor modeling problems.

Firstly we should clarify a reasonable form of separability assumptions about any target tensor. As discussed earlier, without further information, one in general expects the interplay between two parts becomes more complicated as the size $m$ of two parts (or the smaller one, which determines the matricization rank) increases. The complexity of the interplay could be captured by the number of Schmidt components, \textbf{the separability assumption could thus be represented by the maximum number of Schmidt components $N(m)$ as a monotonically increasing function of $m$.}

Now we derive the required universal virtual dimension $\chi$ in different models given $N(m)$. For both HT and TT models, according to Eq.\eqref{minmax}, we have:
\begin{align}
    N(m) = 
    \chi_{{}_{HT, TT}}, \quad \forall m\in [1, L/2].
\end{align}
As the above equation should hold for any value of $m$, and recalling the monotonically increasing nature of $N(m)$, we thus have the final result:
\begin{align}\label{chi_ttht}
    \chi_{{}_{HT, TT}} = N\bigg(\frac{L}{2}\bigg),
\end{align}
This in general is a large number, due to the weakness in the structure of both TT and HT models: there exist single cuts that bipartite two parts with large dimensions.

On the other hand, the situation in MERA is improved as:
\begin{align}
    N(m) = 
    \chi_{{}_{MERA}}^{\log_2{m}}, \quad \forall m\in (1, L/2],
\end{align}
which eventually requires a universal virtual dimension:
\begin{align}\label{chi_mera}
    \chi_{{}_{MERA}} = \max_{m\in(1, \frac{L}{2}]}\big[N(m)\big]^{\frac{1}{\log_2{m}}}.
\end{align}
It is obvious that Eq.\eqref{mera_max} is a special case of the above expression when $N(m)$ is an exponential function $D^m$, i.e. the irreducible problems. Comparing Eq.\eqref{chi_ttht} and Eq.\eqref{chi_mera} by taking the (base-2) logarithm on both expressions, we have:
\begin{align}
    \log_2{\chi_{{}_{MERA}}} &= \max_{m\in(1, \frac{L}{2}]}\log{\bigg(\big[N(m)\big]^{\frac{1}{\log_2{m}}}\bigg)} \nonumber \\
    &= \max_{m\in(1, \frac{L}{2}]}\frac{\log_2{N(m)}}{\log_2{m}} \nonumber \\
    &\leq \max_{m\in(1, \frac{L}{2}]}\log_2{N(m)} \nonumber \\
    &= \log_2{N\bigg(\frac{L}{2}\bigg)}\nonumber \\
    &= \log_2{\chi_{{}_{HT, TT}}},
\end{align}
where the inequality has taken the fact that $m>1$ in general, as the interplay between a single mode and other parts should rarely be the most complicated one.
The above inequality demonstrates the advantage of MERA compared with HT and TT structures.

\section{Discussion}
We discuss the problem of tensor model capacity, and clarifies the difference between tensor complexity and model capacity.
Importantly, a tensor with large complexity does not guarantee sufficient model capacity, if the structure of it implies a strong separability assumption on the targeted problem.
And a Cannikin's law of modeling is proposed, which states that in the scenario of black-box modeling, to ensure a full description of the real world mechanism, the weakest interaction in the model should be stronger than the most complicated interaction in the task.
The concept of entanglement is introduced to the discussion of tensor analysis, which establishes a natural connection between quantum information and tensor analysis.
Based on the proposed separability criteria, new tensor models might be developed accordingly in future studies.

\bibliography{bibfile}

\providecommand{\noopsort}[1]{}\providecommand{\singleletter}[1]{#1}%
\begin{thebibliography}{3}%
\makeatletter
\providecommand \@ifxundefined [1]{%
 \@ifx{#1\undefined}
}%
\providecommand \@ifnum [1]{%
 \ifnum #1\expandafter \@firstoftwo
 \else \expandafter \@secondoftwo
 \fi
}%
\providecommand \@ifx [1]{%
 \ifx #1\expandafter \@firstoftwo
 \else \expandafter \@secondoftwo
 \fi
}%
\providecommand \natexlab [1]{#1}%
\providecommand \enquote  [1]{``#1''}%
\providecommand \bibnamefont  [1]{#1}%
\providecommand \bibfnamefont [1]{#1}%
\providecommand \citenamefont [1]{#1}%
\providecommand \href@noop [0]{\@secondoftwo}%
\providecommand \href [0]{\begingroup \@sanitize@url \@href}%
\providecommand \@href[1]{\@@startlink{#1}\@@href}%
\providecommand \@@href[1]{\endgroup#1\@@endlink}%
\providecommand \@sanitize@url [0]{\catcode `\\12\catcode `\$12\catcode
  `\&12\catcode `\#12\catcode `\^12\catcode `\_12\catcode `\%12\relax}%
\providecommand \@@startlink[1]{}%
\providecommand \@@endlink[0]{}%
\providecommand \url  [0]{\begingroup\@sanitize@url \@url }%
\providecommand \@url [1]{\endgroup\@href {#1}{\urlprefix }}%
\providecommand \urlprefix  [0]{URL }%
\providecommand \Eprint [0]{\href }%
\providecommand \doibase [0]{http://dx.doi.org/}%
\providecommand \selectlanguage [0]{\@gobble}%
\providecommand \bibinfo  [0]{\@secondoftwo}%
\providecommand \bibfield  [0]{\@secondoftwo}%
\providecommand \translation [1]{[#1]}%
\providecommand \BibitemOpen [0]{}%
\providecommand \bibitemStop [0]{}%
\providecommand \bibitemNoStop [0]{.\EOS\space}%
\providecommand \EOS [0]{\spacefactor3000\relax}%
\providecommand \BibitemShut  [1]{\csname bibitem#1\endcsname}%
\let\auto@bib@innerbib\@empty
\bibitem [{\citenamefont {Yu}\ \emph {et~al.}(2017)\citenamefont {Yu},
  \citenamefont {Zheng}, \citenamefont {Anandkumar},\ and\ \citenamefont
  {Yue}}]{yu}%
  \BibitemOpen
  \bibfield  {author} {\bibinfo {author} {\bibfnamefont {R.}~\bibnamefont
  {Yu}}, \bibinfo {author} {\bibfnamefont {S.}~\bibnamefont {Zheng}}, \bibinfo
  {author} {\bibfnamefont {A.}~\bibnamefont {Anandkumar}}, \ and\ \bibinfo
  {author} {\bibfnamefont {Y.}~\bibnamefont {Yue}},\ }\href {\doibase
  10.48550/ARXIV.1711.00073} {\enquote {\bibinfo {title} {Long-term forecasting
  using higher order tensor rnns},}\ } (\bibinfo {year} {2017})\BibitemShut
  {NoStop}%
\bibitem [{\citenamefont {Evenbly}\ and\ \citenamefont
  {Vidal}(2008)}]{evenbly}%
  \BibitemOpen
  \bibfield  {author} {\bibinfo {author} {\bibfnamefont {G.}~\bibnamefont
  {Evenbly}}\ and\ \bibinfo {author} {\bibfnamefont {G.}~\bibnamefont
  {Vidal}},\ }\href {\doibase 10.48550/ARXIV.0801.2449} {\enquote {\bibinfo
  {title} {Entanglement renormalization in free bosonic systems: real-space
  versus momentum-space renormalization group transforms},}\ } (\bibinfo {year}
  {2008})\BibitemShut {NoStop}%
\bibitem [{\citenamefont {Zhang}(2017)}]{yahui}%
  \BibitemOpen
  \bibfield  {author} {\bibinfo {author} {\bibfnamefont {Y.-H.}\ \bibnamefont
  {Zhang}},\ }\href {\doibase 10.48550/ARXIV.1710.05520} {\enquote {\bibinfo
  {title} {Entanglement entropy of target functions for image classification
  and convolutional neural network},}\ } (\bibinfo {year} {2017})\BibitemShut
  {NoStop}%
\end{thebibliography}%

\end{document}